\documentclass[envcountsame]{llncs}

\usepackage{latexsym}
\usepackage{xspace}

\usepackage{amsmath,amssymb}
\usepackage{enumerate}
\usepackage{cite}

\usepackage{booktabs}

\usepackage{url}

\usepackage{algorithm}
\usepackage{algpseudocode}

\newenvironment{proofsketch}{\paragraph{Proof sketch}}{\qedhere\medskip}

\newenvironment{myparagraph}[1]{\medskip\par\noindent\textbf{#1}~~}{}

\def\vS{{\textbf{S}}}

\def\vs{{\textbf{s}}}

\newcommand{\qedhere}{\hspace*{\fill}$\Box$}
\usepackage{marginnote}
\usepackage{alltt}
\usepackage{eepic}
\usepackage{epic}
\usepackage{latexsym}
\usepackage{amssymb}
\usepackage{color,epsfig}
\usepackage{wrapfig}
\usepackage{float}
\usepackage{cite,centernot}

\def\bs{{\textbf{s}}}
\def\bS{{\textbf{S}}}
\def\bv{{\textbf{v}}}
\def\bu{{\textbf{u}}}
\newcommand{\en}{\textit{en\/}}
\newcommand{\act}{\textit{act\/}}
\newcommand{\Prism}{\textsc{Prism}}
\def\cD{\mathcal{D}}

\def\cM{\mathcal{M}}

\def\ra{\rightarrow}

\begin{document}
\title{Distributed Markov Chains}
%T Determined asynchronous probabilistic systems ?
\author{Sumit Kumar Jha\inst{1} \and Madhavan Mukund\inst{2} \and
  Ratul Saha\inst{3} \and P. S. Thiagarajan\inst{3}}
\institute{University of Central Florida, USA,
  \email{jha@eecs.ucf.edu} \and
  Chennai Mathematical Institute, India,
  \email{madhavan@cmi.ac.in} \and
  National University of Singapore, Singapore,
  \email{\{ratul,thiagu\}@comp.nus.edu.sg}
}

\maketitle
\begin{abstract}
  The formal verification of large probabilistic models is important
  and challenging. Exploiting the concurrency that is often present is
  one way to address this problem. Here we study a restricted class of
  asynchronous distributed probabilistic systems in which the
  synchronizations determine the probability distribution for the next
  moves of the participating agents. The key restriction we impose is
  that the synchronizations are \emph{deterministic}, in the sense
  that any two simultaneously enabled synchronizations must involve
  disjoint sets of agents. As a result, this network of agents can be
  viewed as a succinct and distributed presentation of a large global
  Markov chain.  A rich class of Markov chains can be represented this
  way.

  \hspace*{5mm} We define an interleaved semantics for our model in
  terms of the local synchronization actions. The network structure
  induces an independence relation on these actions, which, in turn,
  induces an equivalence relation over the interleaved runs in the
  usual way. We construct a natural probability measure over these
  equivalence classes of runs by exploiting Mazurkiewicz trace theory
  and the probability measure space of the associated global Markov
  chain.

  \hspace*{5mm} It turns out that verification of our model, called
  DMCs (distributed Markov chains), can often be efficiently carried
  out by exploiting the partial order nature of the interleaved
  semantics. To demonstrate this, we develop a statistical model
  checking (SMC) procedure and use it to verify two large distributed
  probabilistic networks.
\end{abstract}

\section{Introduction}

We present here a class of distributed probabilistic systems called
distributed Markov chains (DMCs). A DMC is a network of probabilistic
transition systems that synchronize on common actions. The
synchronzations are deterministic in the sense that two simultaneously
enabled synchronization actions must involve disjoint sets of agents.
The information that the agents gain through a synchronization
determines the probability distribution for their next moves. Internal
actions correspond to synchronizations involving only one agent.

In many distributed probabilistic systems, the communication protocol
can be designed to be deterministic---some examples are discussed in
Section~\ref{section:results}. Hence, the determinacy restriction is
less limiting than may appear at first sight.

We define an interleaved semantics where one synchronization action is
executed at a time. The resulting object is, in general, \emph{not} a
Markov chain. Thus, defining a valid probability measure over
interleaved runs---called \emph{trajectories}--- is a technical
challenge. We address this by noting that there is a natural
independence relation on local actions---two actions are independent
if they involve disjoint sets of agents. Using this relation, we
partition the trajectories into equivalence classes. As usual, each
equivalence class will correspond to a partially ordered
execution. This leads to a trajectory space that resembles the usual
path space of a Markov chain~\cite{baierbook}, except that it will not
be tree-like.  Hence, one cannot readily define a probability measure
over this space.

Due to the determinacy restriction, at any global state any two
enabled actions will be independent. Thus, by letting all the enabled
actions at a global state occur as a single step followed by probabilistic
moves by all the involved agents, one obtains a Markov chain.

Using Mazurkiewicz trace theory~\cite{tracebook}, we then embed the
trajectory space derived from the interleaved semantics into the path
space of the global Markov chain.  This induces a probability measure
over the trajectory space. This is the key technical contribution of
the paper.  We are not aware of a similar result for any well-defined
class of distributed probabilistic systems~\cite{albert1, albert2,
  winskel,DBLP:journals/mst/JesiPS96}.

%T changed slightly

Due to its exponential size (in the number of agents), it will often
be infeasible to analyze a DMC in terms of its global Markov chain. In
contrast, due to the partial order nature of the trajectory space, the
global behaviour of the network can often be efficiently analyzed
using the interleaved semantics. To bring this out, we formulate a
statistical model checking (SMC) problem for DMCs in which the
specifications consist of boolean combinations of local bounded linear
temporal logic (BLTL) \cite{baierbook} formulas. We then develop a
sequential probability ratio test (SPRT) based SMC
procedure~\cite{younesthesis,Wald:1947} to solve this problem.  We
view our SMC procedure as a first step. Other partial order based
reduction techniques such as ample sets \cite{clarkebook} and finite
prefixes of unfoldings \cite{esparzabook} can also be readily
developed for DMCs. Furthermore, one can develop model checking
procedures for more powerful specification logics.

We illustrate the potential of our approach by using our SMC procedure
to analyze two distributed probabilistic algorithms.  The first is a
distributed leader election protocol in an anonymous
ring~\cite{itairodeh}. The second one is a randomized solution to the
classical dining philosophers problem~\cite{LR81}.  We show that for
DMCs, simulations based on asynchronous trajectories are significantly
faster than working directly with the global state space.
%T suppressed the reference to PRISM. I think it will be better to avoid giving 
% the impression we are better than PRISM (though we are; I think PRISM will never % scale) but rather local is better than global for DMCs.

To summarize, our main contribution is identifying determinacy of
communications as a fruitful restriction for distributed stochastic
systems and constructing a probability measure over a partially ordered space of runs. We believe DMCs represent a restricted but clean combination of
concurrent and stochastic dynamics and can lead to
fruitful applications in domains such as embedded
systems~\cite{legay1, viswanathan1,smolka1}, biological
processes~\cite{DBLP:conf/wabi/LangmeadJ07,clarke}, and distributed
protocols~\cite{david2011time, jha2009statistical}.
%T changed a bit
\myparagraph{Related work} Our work is in line with partial order
based methods for Markov Decision Processes (MDPs)~\cite{baier_survey}
where, typically, a partial commutation structure is imposed on the
actions of a \emph{global} MDP.  For instance, in \cite{hermanns},
partial order reduction is used to identify ``spurious''
nondeterminism arising out of the interleaving of concurrent actions,
in order to determine when the underlying behaviour corresponds to a
Markov chain. In contrast, in a DMC, deterministic communication
ensures that local behaviours \emph{always} generate a global Markov
chain.  The independence of actions is directly given by the local
state spaces of the components. This also makes it easier to model how
components influence each other through communications.

The interplay between concurrency and stochasticity has also been
explored in the setting of event structures \cite{albert1, winskel}.
In these approaches, the global behaviour---which is \emph{not} a
Markov chain---is endowed with a probability measure. Further,
probabilistic verification problems are not formulated and
studied. Markov nets, studied in \cite{albert2} can be easily modeled
as DMCs. In \cite{albert2}, the focus is on working out a
probabilistic event structure semantics rather than developing a model
checking procedure based on the interleaved semantics, as we do here.

Our model is formulated as a sub-class of probabilistic asynchronous
automata~\cite{DBLP:journals/mst/JesiPS96}, where we require
synchronizations to be deterministic.  This restriction allows us to
develop a probability measure over the (infinite) trajectory space,
which in turn paves the way for carrying out formal verification based
on probabilistic temporal logic specifications. In contrast, the work
reported in \cite{DBLP:journals/mst/JesiPS96} is language-theoretic,
with the goal of generalizing Zielonka's theorem
\cite{DBLP:journals/ita/Zielonka87} to a probabilistic setting.
However, in the model of \cite{DBLP:journals/mst/JesiPS96},
conflicting actions may be enabled at a global state and it is
difficult to see how one can formulate a $\sigma$-algebra over the
runs with a well-defined probability measure.

\section{The Distributed Markov Chain (DMC) Model}
\label{section:model}

We fix $n$ agents $\{1,2,\ldots,n\}$ and set $[n] = \{1, 2, \ldots,
n\}$. For convenience, we denote various $[n]$-indexed sets of the
form $\{X_i\}_{i \in [n]}$ as just $\{X_i\}$. We begin with some
notation for distributed state spaces.

\begin{definition}
  \label{definition:state-spaces}
  For $i \in [n]$, let $S_i$ be a finite set of local states, where
  $\{S_i\}$ is pairwise disjoint.

  \begin{itemize}

  \item We call $S = \bigcup_i S_i$ the set of \emph{local states}.

  \item For nonempty $u \subseteq [n]$, $\bS_u = \prod_{i \in u}
    S_i$ is the set of $u$-states.

  \item $\bS_{[n]}$ is the set of \emph{global states}, which we
    typically denote $\bS$.

  \item For a state $\bv \in \bS_u$ and $w \subseteq u$, $\bv_w$
    denotes the projection of $\bv$ to $\bS_w$.

  \item For $u = \{i\}$, we write $\bS_i$ and $\bv_i$ rather than
    $\bS_{\{i\}}$ and $\bv_{\{i\}}$, respectively.

  \end{itemize}

\end{definition}

\noindent
Our model is a restricted version of probabilistic
asynchronous automata \cite{DBLP:journals/mst/JesiPS96}.

\begin{definition}
  \label{definition:PAA}
  A \emph{probabilistic asynchronous system} is a structure\linebreak
  $(\{S_i\}, \{s^{in}_i\}, A, loc, en, \{\pi^a\}_{a \in A})$ where:
%T changed "automaton' to "system" since we don't consider final states; can also consider "transition system"
  \begin{itemize}

  \item $S_i$ is a finite set of local states for each $i$ and
    $\{S_i\}$ is pairwise disjoint.

  \item $s^{in}_i \in S_i$ is the initial state of agent $i$.

  \item $A$ is a set of synchronization actions.

  \item $loc: A \ra 2^{[n]} \setminus \emptyset$ specifies the agents
    that participate in each action $a$.

    %T Don't seem to need this immediately.
    % As usual, for $X
    % \subseteq A$, $loc(X) = \bigcup_{a \in X} loc(a)$.

    \begin{itemize}

    \item For $a \in A$, we write $\bS_a$ instead of $\bS_{loc(a)}$ and
      call it the set of $a$-states.

    \end{itemize}

  \item For each $a \in A$, $\en_a \subseteq \bS_a$ is the subset of
    $a$-states where $a$ is enabled.

  \item With each $a \in A$, we associate a probabilistic transition
    function $\pi^a: \en_a \to (\bS_a \to [0,1])$ such that, for every
    $\bv \in \en_a$, $\sum_{\bu \in \bS_a} \pi^a(\bv)(\bu) = 1$.
%TT strengthened a bit.
\end{itemize}
\end{definition}

The action $a$ represents a synchronized communication between the
agents in $loc(a)$ and it is enabled at the global state
$\bs$ if $\bs_a \in \en_a$.  When $a$ occurs at $\bs$, only the
components in $loc(a)$ are involved in the move to the new global
state $\bs'$; the new $a$-state $\bs'_a$ is chosen probabilistically
according to the distribution assigned to the current
$a$-state $\bs_a$ by $\pi^a$.  For every $j \notin loc(a)$, $\bs_j =
\bs'_j$.

We would like to lift the probabilities associated with individual
moves to a probability measure over runs of the system.  This is
difficult to achieve, in general, because of the combination of
nondeterminism, concurrency and probability in the model.  This
motivates us to restrict the nondeterminism in the model.

For an agent $i$ and a local state $s \in S_i$, we define the set of
actions compatible with $s$ to be $\act(s) = \{ a \mid i \in loc(a), s
= \bv_i \mbox{~for some~} \bv \in \en_a\}$.

\begin{definition}
  \label{defn:DMC}
  A \emph{distributed Markov chain (DMC)} is a probabilistic
  asynchronous system $\cD = (\{S_i\}, \{s^{in}_i\}, A, loc, en,
  \{\pi^a\}_{a \in A})$ in which, for each agent $i$ and each local
  state $s \in S_i$, $|\act(s)| = 1$.
\end{definition}

In other words, in a DMC, the set of partners that
an agent can communicate with next is fixed deterministically by its
current local state.  Hence, if two actions $a$ and $b$ are enabled at
%TT shortened to reduce redundancy
a global state $\vs$, they must involve disjoint sets of agents---that
is, $loc(a) \cap loc(b) = \emptyset$.  This allows us to capture the
global behaviour of the model as a Markov chain---as shown in
Section~\ref{section:markov}---whence the name DMC.

%T not needed here; may be when we present the example. But perhaps not even then. %T Often the probability distribution $\pi^a$ associated with an action
%T $a$ can be factored into a product of local distributions $\prod_{i
%T   \in loc(a)} \delta_i$ where $\delta_i: S_i \to [0, 1]$. Thus it will
%T be the case that $\pi^a(\bv)(\bv') = \prod_{i \in loc(a)}
%T \delta_i(\bv_i)$ for every $\bv \in en_a$. This will become clearer
%T when we encounter the example depicted in Fig.~\ref{figure:DMC}.

\subsubsection*{Events}

Let $\cD$ be a DMC.  An \emph{event} of $\cD$ is a triple $e = (\bv,
a, \bv')$ where $\bv,\bv' \in \bS_a$, $\bv \in en_a$ and
$\pi^a(\bv)(\bv') > 0$.  We extend $loc$ to events via
$loc((\bv,a,\bv')) = loc(a)$.

Suppose $e = (\bv, a, \bv')$ is an event and $p =
\pi^a(\bv)(\bv')$. Then $e$ represents an occurrence of the
synchronization action $a$ followed by a joint move by the agents in
$loc(a)$ from $\bv$ to $\bv'$ with probability $p$.  Again,
components outside $loc(e)$ are unaffected by this move.  

%We call
%$\bv$ the set of \emph{pre-conditions} and $\bv'$ the set of
%\emph{post-conditions} for $e$.
%TT I don't think we use this terminology anywhere.
Let $\Sigma$ denote the set of events of $\cD$ and $e, e',\ldots$
range over $\Sigma$. With the event $e = (\bv, a, \bv')$ we
associate the probability $p_e = \pi^a(\bv)(\bv')$.

\subsubsection*{The interleaved semantics}

We now associate a global transition system with $\cD$ based on event
occurrences.

Recall that $\bS$ is the set of global states.  The event $e = (\bv,
a, \bv')$ is \emph{enabled} at $\vs \in \bS$ iff $\bv = \vs_a \in en_a$.
The transition system of $\cD$ is $TS =
(\vS,\Sigma,\to,\vs^{in})$, where ${\to} \subseteq \vS \times (\Sigma
\times (0, 1]) \times \vS$ is given by $\vs \xrightarrow{e, p_e} \vs'$
iff $e = (\bv, a , \bv')$ is enabled at $\vs$, $\bs'_a = \bv'$ and
$\vs_j = \vs'_j$ for every $j \notin loc(e)$.

\begin{figure}[t]
  \vspace*{-2ex}
  \hspace*{\fill}
    \includegraphics[width=0.9\textwidth]{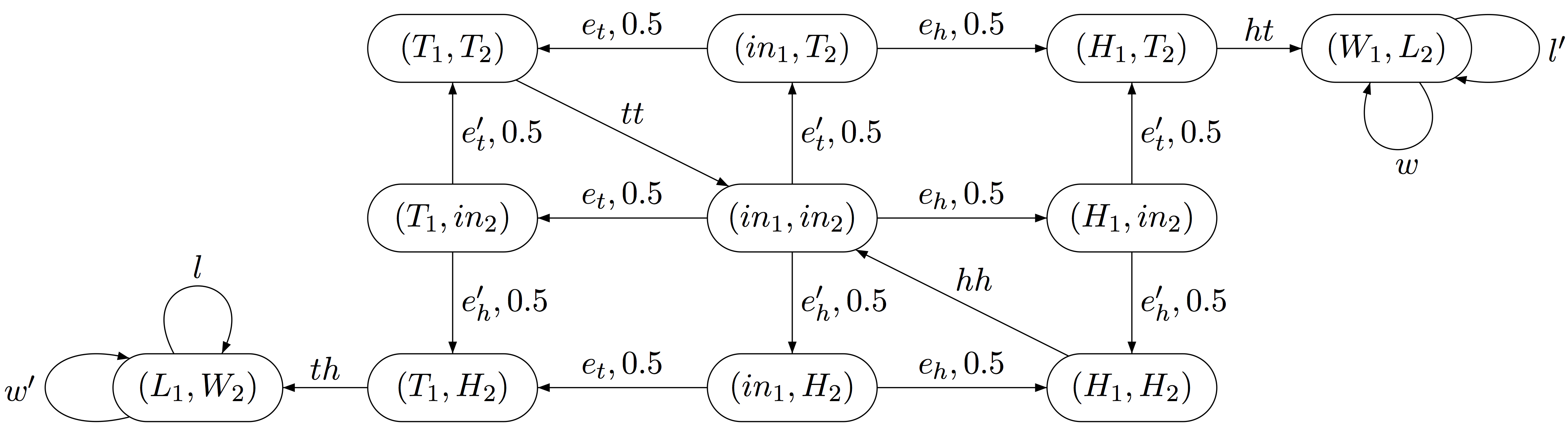}
  \hspace*{\fill}
  \vspace*{-1ex}
  \caption{The transition system of a DMDP for a two player game}
  \label{figure:DMC}
  \vspace*{-3ex}
\end{figure}

In Fig.~\ref{figure:DMC} we show the transition system of a DMC
describing a simple two player game.  Each player tosses an unbiased
coin. If the tosses have the same outcome the players toss again.  If
the outcomes are different then the player who tossed heads wins. In
this $2$-component system, $S_i = \{in_i, T_i, H_i, L_i, W_i\}$ for $i
= 1, 2$, where $T_i/H_i$ denote that a tail/head was tossed,
respectively, and $L_i/W_i$ denote local losing/winning states,
respectively.  Agent $1$, for instance, has an internal action $a_1$
with $loc(a_1) = \{1\}$, $en_{a_1} = \{in_1\}$ and
$\pi^{a_1}(in_1)(T_1) = 0.5 = \pi^{a_1} (in_1)(H_1)$. Thus $e_h =
(\{in_1\}, a_1, \{H_1\})$ and $e_t = (\{in_1\}, a_1, \{T_1\})$ are
both events that are enabled at $(in_1, in_2)$.  Symmetrically, agent
$2$ has the internal action $\{a_2\}$ with $loc(a_2) = \{2\}$,
$en_{a_2} = \{in_2\}$ and corresponding events $e'_h$ and $e'_t$.  On
the other hand, $b$ is an action with $loc(b) = \{1, 2\}$, $en_b =
\{(T_1,T_2)\}$ where $tt = (\{T_1, T_2\}, b, \{in_1, in_2\})$ is an
event with $\pi^b((T_1, T_2))((in_1, in_2)) = 1$. To aid readability,
if the probability of an event is $1$ then this value is not shown.

\myparagraph{The trace alphabet $(\Sigma, I)$} We conclude this
section by defining the independence relation $I \subseteq \Sigma
\times \Sigma$ given by $e~I~e'$ iff $loc(e) \cap loc(e') =
\emptyset$. Clearly $I$ is irreflexive and symmetric and hence
$(\Sigma, I)$ is a Mazurkiewicz trace alphabet~\cite{tracebook}.

\section{The trajectory space}
\label{section:trajectory}

Let $TS$ be the transition system associated with a DMC $\cD$.  To
reason about the probabilistic behaviour of $TS$, one must follow the
technique used for Markov chains---namely, build a $\sigma$-algebra
over the paths of this transition system, endowed with a probability
measure. The major difficulty is that, due to the mix of concurrency
and stochasticity, $TS$ is not a Markov chain in general. In
Fig.~\ref{figure:DMC}, for instance, the sum of the probabilities of
the transitions originating from the state $(in_1, in_2)$ is $2$.  To
get around this, we will filter out concurrency by working with
equivalence classes of paths rather than individual paths.

We refer to paths in $TS$ as \emph{trajectories}.  A finite
\emph{trajectory} of $TS$ from $\vs \in \vS$ is a sequence of the form
$\vs_0 e_0 \vs_1$ $\ldots$ $\vs_{k-1} e_{k-1} \vs_k$ such that $\vs_0
= \vs$ and, for $0 \leq \ell < k$, $\vs_{\ell} \xrightarrow{e_{\ell},
  p_{\ell}} \vs_{\ell{+1}}$ (with $p_{\ell} = p_{e_{\ell}}$).
Infinite trajectories are defined as usual.

For the trajectory $\rho = \vs_0 e_0 \vs_1 \ldots \vs_{k-1} e_{k-1}
\vs_k$, we define $ev(\rho)$ to be the event sequence $e_0 e_1 \ldots
e_{k-1}$. Again, this notation is extended to infinite trajectories in
the natural way. Due to concurrency, one can have infinite
trajectories that are not maximal, so we proceed as follows.

Let $\Sigma_i = \{e \mid i \in loc(e)\}$. Suppose $\xi$ is an event
sequence (finite or infinite).  Then $proj_i(\xi)$ is the sequence
obtained by erasing from $\xi$ all events that are not in
$\Sigma_i$. This leads to the equivalence relation $\approx$ over
event sequences given by $\xi \approx \xi'$ iff $proj_i(\xi) =
proj_i(\xi')$ for every $i$. We let $[\xi]$ denote the
$\approx$-equivalence class containing $\xi$ and call it a
\emph{(Mazurkiewicz) trace}.\footnote{For infinite sequences, it is
  technically more convenient to define traces using projection
  equivalence rather than permutation of independent actions.}  The
partial order relation $\sqsubseteq$ over traces is defined as $[\xi]
\sqsubseteq [\xi']$ iff $proj_i(\xi)$ is a prefix of $proj_i(\xi')$
for every $i$. Finally the trace $[\xi]$ is said to be maximal iff for
every $\xi'$, $[\xi] \sqsubseteq [\xi']$ implies $[\xi] = [\xi']$.
The trajectory $\rho$ is \emph{maximal} iff $[ev(\rho)]$ is a maximal
trace. In the transition system of Fig.~\ref{figure:DMC}, $(in_1,
in_2) e_h (H_1, in_2) e'_T (H_1, T_2) ht ((W_1, L_2) l')^{\omega}$ is
a non-maximal infinite trajectory.

\vspace*{-1ex}

\subsubsection*{The $\sigma$-algebra of trajectories}

We denote by $Trj_{\vs}$ the set of maximal trajectories from
$\vs$. Two trajectories can correspond to interleavings of the same
partially ordered execution of events. Hence one must work with
equivalence classes of maximal trajectories to construct a probability
measure.  The equivalence relation $\simeq$ over $Trj_{\vs}$ that we
need is defined as $\rho \simeq \rho'$ iff $ev(\rho) \approx
ev(\rho')$. As usual $[\rho]$ will denote the equivalence class
containing the trajectory $\rho$.
%T added a sentence.

%M Omit the following paragraph?
%Thus, equivalence classes of maximal trajectories will correspond to
%infinite paths in a Markov chain. Naturally, we will use equivalence
%classes of finite trajectories as the counterpart of finite paths in a
%Markov chain in order to identify the basic cylindrical sets.

Let $\rho$ be finite trajectory from $\vs$. Then ${\uparrow}\rho$ is
the subset of $Trj_{\vs}$ satisfying $\rho' \in {\uparrow}\rho$ iff
$\rho$ is a prefix of $\rho'$. We now define $BC(\rho)$, the basic
$trj$-cylinder at $\vs$ generated by $\rho$, to be the least subset of
$Trj_{\vs}$ that contains ${\uparrow}\rho$ and satisfies the closure
property that if $\rho' \in BC(\rho)$ and $\rho' \simeq \rho''$ then
$\rho'' \in BC(\rho)$.  In other words, $BC(\rho) = \{ [\rho'] \mid
\rho' \in Trj_{\vs}, [ev(\rho)] \sqsubseteq [ev(\rho')] \}$.

It is worth noting that we could have $BC(\rho) \cap BC(\rho') \neq
\emptyset$ without having $\rho \simeq \rho'$. For instance, in
Fig.~\ref{figure:DMC}, let $\rho = (in_1, in_2) e_h (H_1, in_2)$ and
$\rho' = (in_1, in_2) e'_t (in_1, T_2)$. Then $BC(\rho)$ and
$BC(\rho')$ will have common maximal trajectories of the form $(in_1,
in_2) e_h (H_1, in_2) e'_t (H_1, T_2) \ldots$.

We now define $\widehat{SA}(\vs)$ to be the least $\sigma$-algebra
that contains the basic $trj$-cylinders at $\vs$ and is
closed under countable unions and complementation (relative to
$Trj_{\vs}$).

To construct the probability measure $\widehat{P}: \widehat{SA}(\vs)
\to [0, 1]$ we are after, a natural idea would be to assign a
probability to each basic $trj$-cylinder as follows.  Let $BC(\rho)$
be a basic $trj$-cylinder with $\rho = \vs_0 e_0 \vs_1 \ldots
\vs_{k-1} e_{k-1} \vs_k$.  Then $\widehat{P}(BC(\rho)) = p_0 \cdot p_1
\ldots p_{k-1}$, where $p_{\ell} = p_{e_{\ell}}$, for $0 \leq \ell <
k$.  This is inspired by the Markov chain case in which the
probability of a basic cylinder is defined to be the product of the
probabilities of the events encountered along the common finite prefix
of the basic cylinder. However, showing directly that this extends
canonically to a probability measure over $\widehat{SA}_{\vs}$ is very
difficult.

We shall tackle this problem by associating a Markov chain $\cM$ with $\cD$ and then embedding  $\widehat{SA}_{\vs}$ into
$SA_{\vs}$, the $\sigma$-algebra generated by the infinite paths in
$\cM$ starting from $\vs$. The standard probability
measure over $SA_{\vs}$ will then  induce a probability measure over
$\widehat{SA}_{\vs}$.

\section{The Markov chain semantics}
\label{section:markov}

We associate a Markov chain with  a DMC using the notion of
maximal steps with respect to the trace alphabet $(\Sigma,I)$ . A
\emph{nonempty} set of events $u \subseteq \Sigma$ is a \emph{step} at
$\vs$ iff each $e \in u$ is enabled at $\vs$ and for every distinct
pair of events $e, e'\in u$, $e~I~e'$. We say $u$ is a \emph{maximal}
step at $\vs$ iff $u$ is a step at $\vs$ and $u \cup \{e\}$ is not a
step at $\vs$ for any $e \notin u$.  In Fig.~\ref{figure:DMC},
$\{e_h, e'_h\}$, $\{e_h, e'_t\}$, $\{e_t, e'_h\}$ and $\{e_t, e'_t\}$
are maximal steps at the initial state $(in_1, in_2)$.

Let $u$ be a maximal step at $\vs$. Then $\vs'$ is the $u$-successor
of $\vs$ iff the following conditions are satisfied: \textbf{(i)}~For
each $e \in u$, if $e = (\bv, a, \bv')$ and $i \in loc(e)$ then
$\vs'_i = \bv'_i$, and \textbf{(ii)}~$\vs_j = \vs'_j$ if $j \notin
loc(u)$, where, as usual, $loc(u) = \bigcup_{e \in u} loc(e)$.

Suppose $u$ is a maximal step at $\vs$ and $i \in loc(u)$. Then,
because events in a step are independent, it follows that there exists
a unique $e \in u$ such that $i \in loc(e)$, so the $u$-successor of
$\vs$ is unique. We say $\vs'$ is a \emph{successor} of $\vs$ iff
there exists a maximal step $u$ at $\vs$ such that $\vs'$ is the
$u$-successor of $\vs$.  From the definition of a DMC, it is easy to
see that if $\vs'$ is a successor of $\vs$ then there exists a unique
maximal step $u$ at $\vs$ such that $\vs'$ is the $u$-successor of
$\vs$. Finally, we say that $\vs$ is a \emph{deadlock} iff no event is
enabled at $\vs$.

\begin{definition}
  \label{definition:MarkovChain}

  The Markov chain $\cM : \vS \times \vS \to [0, 1]$ generated
  by $\cD$ is given by:

  \begin{itemize}

  \item If $\vs \in \vS$ is a deadlock then $\cM(\vs, \vs) = 1$
    and $\cM(\vs, \vs') = 0$ if $\vs \neq
    \vs'$.

  \item Suppose $\vs \in \vS$ is not a deadlock. Then $\cM(\vs,
    \vs') = p$ iff there exists a maximal step $u$ at $\vs$ such that
    $\vs'$ is the $u$-successor of $\vs$ and $p = \prod_{e \in u}
    p_e$.

  \item If $\vs$ is not a deadlock and $\vs'$ is not a successor of
    $\vs$ then $\cM(\vs, \vs') = 0$.

  \end{itemize}
\end{definition}

It follows that $\cM(\vs, \vs') \in [0,1]$ for every $\vs, \vs' \in
\vS$.  In addition, if $u$ and $u'$ are two maximal steps at $\vs$
then $loc(u) = loc(u')$ and $|u| = |u'|$. Using these facts it is easy
to verify that $\cM$ is indeed a finite state Markov chain. The
initial state of $\cM$ is $\vs^{in} = (s^{in}_1,
s^{in}_2,\ldots,s^{in}_n)$.

% Madhavan: Moved figure to top of page
\begin{figure}[t]
  \vspace*{-3ex}
  \hspace*{\fill}
    \includegraphics[width=0.8\textwidth]{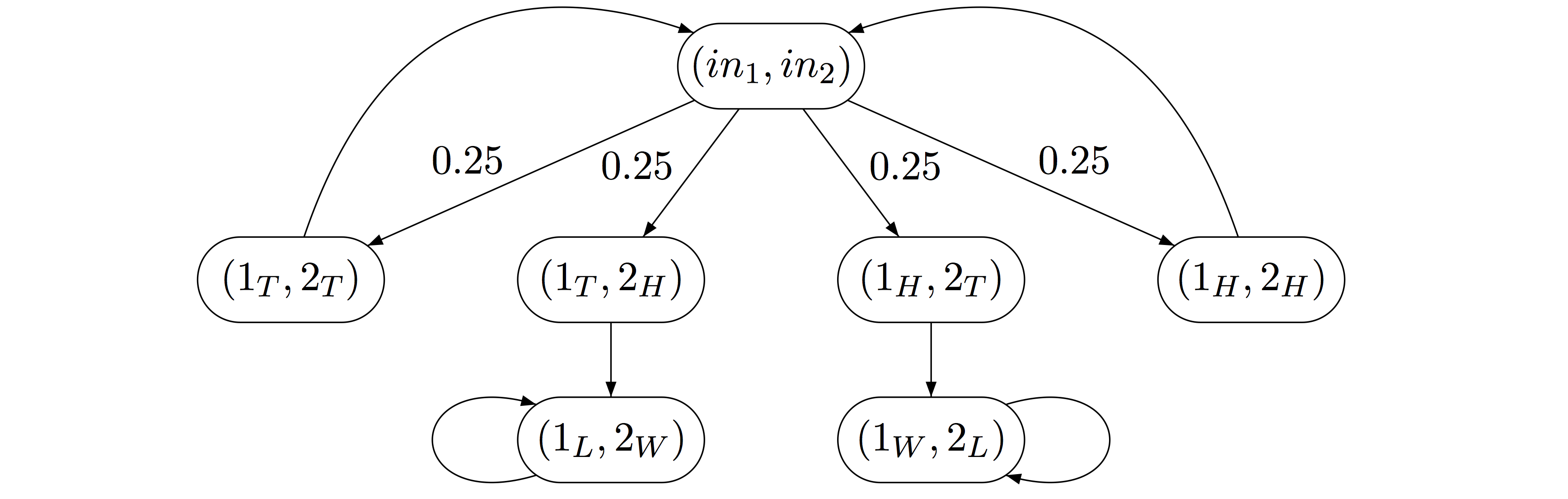}
  \hspace*{\fill}
  \vspace*{-1ex}
  \caption{Markov chain for the DMC in Fig~\ref{figure:DMC}}
  \label{figure:Markov}
  \vspace*{-3ex}
\end{figure}

In Fig.~\ref{figure:Markov} we show the Markov chain of the DMC whose
transition system was shown in Fig.~\ref{figure:DMC}.  Again,
unlabelled transitions have probability $1$. A DMC may have a
\emph{reachable} deadlock---a sequence of global states $\vs_0 \vs_1
\ldots \vs_k$ such that $\vs_0 = \vs^{in}$, $M(\vs_{\ell}, \vs_{\ell +
  1}) > 0$, for $0 \leq \ell < k$, and $\vs_k$ is a deadlock.  One can
effectively check for reachable deadlocks by forming a
\emph{non-deterministic} asynchronous transition system and analyze it
using traditional verification methods. Henceforth,
for simplicity, we assume that DMCs are free of (reachable)
deadlocks. Our results can be easily extended to handle deadlocks.

Suppose $u$ is a maximal step at $\vs$ with $|u| = m$ and $|S_i| = k$
for each $i \in loc(u)$. In $\cM$ there will be, in general, $k^m$
transitions at $\vs$. In contrast there will be at most $k \cdot m$
transitions at $\vs$ in $TS$. Hence---assuming that we do not
explicitly construct $\vS$---there can be substantial computational
gains if one can verify the properties of $\cD$ by working with $TS$
instead of $\cM$.  This will become clearer when we look at some
larger examples in Section~\ref{section:results}.

\subsubsection*{The path space of $\cM$}

Let $\cM$ be the Markov chain associated with a DMC $\cD$.
A finite path in $\cM$ from $\vs$ is a sequence $\tau = \vs_0 \vs_1
\ldots \vs_m$ such that $\vs_0 = \vs$ and $\cM(\vs_{\ell}, \vs_{\ell +
  1}) > 0$, for $0 \leq \ell < m$. The notion of an infinite path
starting from $\vs$ is defined as usual. $Path_{\vs}$ and
$Path^{fin}_{\vs}$ denote the set of infinite and finite paths
starting from $\vs$, respectively.

For $\tau \in Paths^{fin}_{\vs}$, ${\uparrow}\tau \subseteq
Paths_{\vs}$ is the set of infinite paths that have $\tau$ as a
prefix.  $\Upsilon \subseteq Path_{\vs}$ is a basic cylinder at $\vs$
if $\Upsilon = {\uparrow}\tau$ for some $\tau \in Paths^{fin}_{\vs}$.
The $\sigma$-algebra over $Path_{\vs}$, denoted $SA(\vs)$, is the
least family that contains the basic cylinders at $\vs$ and is closed
under countable unions and complementation (relative to
$Path_{\vs}$). $P_{\vs} : SA(\vs) \to [0, 1]$
is the usual probability measure that assigns to each basic cylinder
${\uparrow}\tau$, with $\tau = \vs_0 \vs_1 \ldots \vs_m$, the
probability $p = p_0 \cdot p_1 \cdots p_{m{-1}}$, where
$\cM(\vs_{\ell}, \vs_{\ell + 1}) = p_{\ell}$, for $0 \leq \ell < m$.

\section{The probability measure for the trajectory space}
\label{section:relating}

To construct a probability measure over the trajectory space we shall
associate infinite paths in $\cM$ with maximal trajectories in
$TS$. The Foata normal form from Mazurkiewicz trace theory will help
achieve this.  Let $\xi \in \Sigma^{\star}$. A standard fact is that
$[\xi]$ can be canonically represented as a ``step'' sequence of the
form $u_1 u_2 \ldots u_k$.  More precisely, the Foata normal form of
the finite trace $[\xi]$, denoted $FN([\xi])$, is defined as
follows~\cite{tracebook}.

\begin{itemize}

\item $FN([\epsilon]) = \epsilon$.

\item Suppose $\xi = \xi' e$ and $FN([\xi']) = u_1 u_2 \ldots
  u_{k}$. If there exists $e' \in u_{k}$ such that $(e', e) \notin I$ then
  $FN([\rho]) = u_1 u_2 \ldots u_{k} \{e\}$. If not, let $\ell$ be the
  least integer in $\{1, 2, \ldots, k\}$ such that $e~I~e'$ for every
  $e' \in \bigcup_{\ell \leq m \leq k} u_m$. Then $FN([\rho]) = u_1
  \ldots u_{\ell{-1}} (u_{\ell} \cup \{e\}) u_{\ell{+1}} \ldots u_m$.
%T replaced D since it has not been defined.
\end{itemize}

\noindent
For the example shown in Fig.~\ref{figure:DMC},
$FN(e_h~e'_t~ht~\ell'~w~w) = \{e_h, e'_t\}~\{ht\}~\{w,
\ell'\}~\{w\}$. This notion is extended to infinite traces in the
obvious way. Note that $\xi \approx \xi'$ iff $FN(\xi) =FN(\xi')$.

Conversely, we can extract a (maximal) step sequence from a path in
$\cM$.  Suppose $\vs_0 \vs_1 \ldots$ is a path in $Paths_{\vs}$.
There exists a unique sequence $u_1 \, u_2\ldots$ such that $u_{\ell}$
is a maximal step at $\vs_{\ell - 1}$ and $\vs_{\ell}$ is the
$u_{\ell}$-successor of $\vs_{\ell - 1}$ for every $\ell > 0$. We let
$st(\tau) = u_1 \, u_2 \ldots$ and call it the step sequence induced
by $\tau$.

This leads to the map $tp : Trj_{\vs} \to Paths_{\vs}$ given by
$tp(\rho) = \tau$ iff $FN(ev(\rho)) = st(\tau)$. It is easy to check
that $tp$ is well-defined.  As usual, for $X \subseteq Trj_{\vs}$ we
define $tp(X) = \{tp(\rho) \mid \rho \in X\}$. It turns out that $tp$
maps each basic cylinder in the trajectory space to a finite union of
basic cylinders in the path space.  As a result, $tp$ maps every
measurable set of trajectories to a measurable set of
paths. Consequently, one can define the probability of a measurable
set of trajectories $X$ to be the probability of the measurable set of
paths $tp(X)$.

To understand how $tp$ acts on the basic cylinder $BC(\rho)$, let
$FN(ev(\rho)) = u_1 u_2 \ldots u_k$.  We associate with $\rho$ the set
of finite paths $paths(\rho) = \{ \pi \mid st(\pi) = U_1 U_2 \ldots
U_k \mbox{~and~} u_{\ell} \subseteq U_{\ell} \mbox{~for~} 1 \leq \ell
\leq k\}$. In other words $\pi \in paths(\rho)$ if it extends each
step in $FN(ev(\rho))$ to a maximal step.  Then, $tp$ maps $BC(\rho)$
to the (finite) union of the basic cylinders in $paths(\rho)$. These
observations and their main consequence, namely the construction of a
probability measure over the trajectory space, can be summarized as:
%T added this bit. Hopefully the results section can be correspondingly shortened.

\begin{lemma}
  \label{lemma:main}
  \hspace*{\fill}\vspace*{-1.5ex}\par

  \begin{enumerate}[(i)]

  \item Let $B = BC(\rho)$ be a basic $trj$-cylinder from $\vs$,
    with $FN(ev(\rho)) = u_1 u_2 \ldots u_k$. Then $tp(B)$ is a
    finite union of basic cylinder sets in $SA(\vs)$ and is hence a
    member of $SA(\vs)$. Furthermore $P(tp(B)) = \prod_{1 \leq \ell
      \leq k} p_{\ell}$ where $p_{\ell} = \prod_{e \in u_{\ell}}
    p_e$ for $1 \leq \ell \leq k$.

  \item If $B \in \widehat{SA}(\vs)$ then $tp(B) \in SA(\vs)$.

  \item Define $\widehat{P}: \widehat{SA}(\vs) \ra [0, 1]$ as
    $\widehat{P}(B) = P(tp(B))$. Then $\widehat{P}$ is a probability
    measure over $\widehat{SA}(\vs)$.

  \end{enumerate}
\end{lemma}

\begin{proofsketch}
  Let $BC(\rho)$ be the basic $trj$-cylinder from $\vs$ generated by
  $\rho \neq \epsilon$ and $FN(ev(\rho)) = u_1 u_2 \ldots
  u_k$. Suppose $\tau \in Path_{\vs}$.  Then, using the semantic
  definitions, it is tedious but straightforward to show that $\tau
  \in tp(BC(\rho))$ iff $u_i \subseteq st(\tau)(\ell)$, for $1 \leq
  \ell \leq k$. (Here, $st(\tau)(\ell)$ is the maximal step appearing
  in position $\ell$ of the sequence $st(\tau)$.) It will then follow
  that $tp(BC(\rho))$ is a finite union of basic cylinder sets in
  $SA(\vs)$ and is hence a member of $SA(\vs)$. Furthermore, one can
  argue that $P(tp(BC(\rho)) = \prod_{1 \leq \ell \leq k} p_{\ell}$.

  For the other two parts, we first establish easily that if $B \in
  \widehat{SA}(\vs)$, $\rho \in B$ and $\rho \simeq \rho'$ then $\rho'
  \in B$ as well. Next, it is straightforward to show that if $B, B'
  \in \widehat{SA}(\vs)$ with $B \cap B' = \emptyset$ then $tp(B) \cap
  tp(B') = \emptyset$ too.  Finally, one can also show $tp$ is
  onto. Using these facts, the second and third parts of the lemma can
  be easily established.
\end{proofsketch}

% Madhavan: Added this paragraph to clarify what is going on

Note that while a finite path in $\cM$ always induces a maximal step
sequence, a finite trajectory, in general, does not have this
structure. Some components can get ahead of others by an arbitrary
amount. The lemma above states that, despite this, any finite
trajectory defines a basic cylinder whose probability can be easily
computed. This helps considerably when verifying the properties of
$\cM$. In particular local reachability properties can be checked by
exercising only those components that are relevant.

Going back to our running example let $\rho_t = (in_1, in_2) e_t (T_1,
in_2))$, and $X_t = {\uparrow}\rho_t$. Let $\rho'_t = (in_1, in_2) e'_t
(in_1, T_2))$, and $X'_t = {\uparrow}\rho_t$. Assume $\rho_h$, $X_h$,
$\rho'_h$ and $X'_h$ are defined similarly. Then $\widehat{P}(X_t) =
\widehat{P}(X'_h) = 0.5$ while $\widehat{P}(X_h \cup X_t) = 1$. On the
other hand due to the fact that $e_h$ and $e'_h$ are independent we
will have $\widehat{P}(X_h \cup X'_h) = 0.75$.

\section{Model checking $PBLTL^{\otimes}$ specifications}
\label{section:logic}

We have designed a statistical model checking procedure to verify
dynamic properties of DMCs. The specification logic $PBLTL^{\otimes}$
(product $PBLTL$) is a simple generalization of probabilistic bounded
linear time temporal logic ($PBLTL$) \cite{jha} that captures boolean
combinations of local properties of the components.  The logic can
express interesting global reachablity properties as well as the
manner in which the components influence each other.

We assume a collection of pairwise disjoint sets of atomic
propositions $\{AP_i\}$.  As a first step, the formulas of
$BLTL^{\otimes}$ are given as follows.

\begin{enumerate}[(i)]

\item $ap \in AP_i$ is a $BLTL^{\otimes}$ formula and $type(ap) =
  \{i\}$.

\item If $\varphi$ and $\varphi'$ are $BLTL^{\otimes}$ formulas with
  $type(\varphi) = type(\varphi') = \{i\}$ then so is $\varphi
  \textbf{U}^{t}_i \varphi'$ where $t$ is a non-negative
  integer. Further, $type(\varphi \textbf{U}^{ t}_i \varphi') =
  \{i\}$.  As usual, $\Diamond^t \varphi$ abbreviates
  $true~\textbf{U}^t \varphi$ and $\Box^t \varphi$ is defined as
  $\lnot \Diamond^t \lnot \varphi$.

\item If $\varphi$ and $\varphi'$ are $BLTL^{\otimes}$ formulas then
  so are $\lnot \varphi$ and $\varphi \lor \varphi'$ with $type(\lnot
  \varphi) = type(\varphi)$ and $type(\varphi \lor \varphi') =
  type(\varphi) \cup type(\varphi')$.

\end{enumerate}

\noindent
The formulas of $PBLTL^{\otimes}$ are given by:

\begin{enumerate}[(i)]

\item Suppose $\varphi$ is a $BLTL^{\otimes}$ formula and $\gamma$ a
  rational number in the open interval $(0,1)$.  Then $Pr_{\geq
    \gamma}(\varphi)$ is a $PBLTL^{\otimes}$ formula.
  % T changed P to Pr

\item If $\psi$ and $\psi'$ are $PBLTL^{\otimes}$ formulas then so are
  $\lnot \psi$ and $\psi \lor \psi'$.

\end{enumerate}

To define the semantics, we project each trajectory  to its
components.  For $\vs \in \vS$ and $i \in [n]$ we define $Proj_i :
Trj^{fin}_{\vs} \to S_i^{+}$ inductively.

\begin{enumerate}[(i)]

\item $Proj_i(\vs) = \vs_i$.

\item Suppose $\rho = \vs_0 e_o \vs_1 \ldots \vs_m e_m \vs_{m+1}$ is
  in $Trj^{fin}_{\vs}$ and $\rho' = \vs_0 e_0 \vs_1 \ldots \vs_m$.  If
  $i \in loc(e_m)$ then $Proj_i(\rho) = Proj_i(\rho')
  (\vs_{m+1})_i$. Otherwise $Proj_i(\rho) = Proj_i(\rho')$.

\end{enumerate}

\noindent
We lift $Proj_i$ to infinite trajectories in the obvious way---note
that $Proj_i(\rho)$ can be a finite sequence for the infinite
trajectory $\rho$.
%T knocked off a sentence.
We assume a set of local valuation functions $\{V_i\}$, where $V_i:
S_i \to 2^{AP_i}$.
%T edited a bit
Let $\varphi$ be a $BLTL^{\otimes}$ formula with $type(\varphi) =
\{i\}$. We begin by interpreting such formulas over sequences
generated by the alphabet $S_i$. For $\varrho \in S^+_i \cup
S^\omega_i$,
%T inserted the subscript!
the satisfaction relation $\varrho, k \models_i \varphi$, with $0 \leq
k \leq |\varrho|$, is defined as follows.

\begin{enumerate}[(i)]

\item $\varrho, k \models_i ap$ for $ap \in AP_i$ iff $ap \in
  V_i(\varrho(k)(i))$, where $\varrho(k)(i)$ is the $S_i$-state at
  position $k$ of the sequence $\varrho$.

\item $\lnot$ and $\lor$ are interpreted in the usual way.

% Madhavan: Fixed the semantics of bounded until.
%\Comment{It is a bit hidden that projection removes stuttering, so
%  until bound for $i$-type formulas is actual $i$-moves, not
%  global steps.}

\item $\varrho, k \models_i \varphi_1 \textbf{U}^{t}_i \varphi_2$
  iff there exists $\ell$ such that $k \leq \ell \leq max(k + t,
  |\varrho|)$ with $\varrho, \ell \models_i \varphi_2$, and
  $\varrho, m \models_i \varphi_1$, for $k \leq m < \ell$.

\end{enumerate}

\noindent
As usual, $\varrho \models_i \varphi$ iff $\varrho, 0 \models_i
\varphi$. Next, suppose $\varphi$ is a $BLTL^{\otimes}$ formula and
$\rho \in Path_{\vs}$. Then the relation $\rho \models_{\vs} \varphi $
is defined as follows.

\begin{enumerate}[(i)]

\item If $type(\varphi) =\{i\}$ then $\rho \models_{\vs} \varphi $
  iff $Proj_i(\rho) \models_i \varphi$.

\item Again, $\lnot$ and $\lor$ are interpreted in the standard way.

\end{enumerate}

\noindent
Given a formula $\varphi$ in $BLTL^{\otimes}$ and a global state
%T changed PBLTL to BLTL
$\vs$, we define $Trj_{\vs}(\varphi)$ to be the set of trajectories
$\{\rho \in Trj_{\vs} \mid \rho \models_{s} \varphi\}$.

\begin{lemma}
  For every formula $\varphi$, $Trj_{\vs}(\varphi)$ is a member of
  $\widehat{SA}(\vs)$.
\end{lemma}

\begin{proofsketch}
  If we interpret the formulas over $\cM$, we easily derive that
  $Path_{\vs}(\varphi)$ is a member of $SA(\vs)$ for every
  $\varphi$. We then use Lemma~\ref{lemma:main} to obtain this
  result.
\end{proofsketch}

\noindent
The semantics of $PBLTL^{\otimes}$ is now given by the relation $\cD
\models^{trj}_{\vs} \psi$, defined as:
%T knocked off a sentence. changed P_{\vs} below to P. introduced superscript "path"
%M Superscript should be "trj"!
\begin{enumerate}[(i)]

\item Suppose $\psi = Pr_{\geq \gamma}(\varphi)$. Then $\cD
  \models^{trj}_{\vs} \psi$ iff $\widehat{P}(Path_{\vs}(\varphi)) \geq
  \gamma$.

\item Again, the interpretations of $\lnot$ and $\lor$ are the
  standard ones.

\end{enumerate}

% Moved this example after the semantics
\noindent
For the example in Fig.~\ref{figure:DMC}, one can assert
$\widehat{P}_{\geq 0.99}((\Diamond^7 L_1 \land \Diamond^7 W_2) \lor
(\Diamond^7 W_1 \land \Diamond^7 L_2))$. Here the local states also
serve as the atomic propositions.  Hence, the formula says that with
probability $\geq 0.99$, a winner will be decided within $7$ rounds.

%J

We write $\cD \models^{trj} \psi$ for $\cD \models^{trj} _{\vs^{in}}
\psi$. The model checking problem is to determine whether $\cD
\models^{trj} \psi$. We shall adapt the SMC procedure developed in
\cite{younes} to solve this problem approximately.
% Madhavan: knocked of one more sentence
%The first step is to translate our model checking problem
%to the interleaved setting.
%T knocked off a sentence. we have said it before.

\section{Statistical model checking}
\label{section:smc}

Given a DMC $\mathcal{D}$ and a $PBLTL^{\otimes}$ specification
$\psi$, our goal is to determine whether $\cD \models^{trj} \psi$
(that is, $\cD \models^{trj}_{s^{in}} \psi$).  We develop a statistical
%T changed a bit.
model checking (SMC) procedure to provide an approximate solution to
this problem.  We note that in the Markov chain setting, given a BLTL
formula and a path in the chain, there is a bound $k$ that depends
only on the formula such that we can decide whether the path is a
model of the formula by examining just a prefix of length $k$ of the
path~\cite{jha}. By the same reasoning, for a $BLTL^{\otimes}$ formula
$\varphi$, we can compute a vector of bounds $(k_1, k_2, \ldots, k_n)$
that depends only on $\varphi$ such that for any trajectory $\rho$
starting from $\vs^{in}$, we only need to examine a finite prefix
$\rho'$ of $\rho$ that satisfies $|Proj_i(\rho')| \geq k_i$, for $1
\leq i \leq n$. The complication in our setting is that such a prefix
of $\rho$ may not exist.
%T changed \leq to \geq above!

%M reintroduced blank line to separate the paragraphs
To cope with this, we maintain a count vector $(c_1, c_2, \ldots,
c_n)$ that records how many times each component has moved along the
trajectory $\rho$ that has
been generated so far. A simple reachability analysis will reveal
whether a component is \emph{dead} in the current global state; that
is, starting from the current state, there is no possibility of
reaching a state in which an event involving this agent can be
executed. We mark such components as dead. We then execute, one by
one, all the enabled actions---using a fixed linear order over the set
of actions---followed by one move by each of the participating agents
according to the underlying probabilities.  Recall that action $a$ is
enabled at $\vs$ iff $\vs_a \in \en_a$.  Due to
the determinacy of communication, the global state thus reached will
depend only on the probabilistic moves chosen by the participating
agents. We then update the count vector to $(c'_1, c'_2, \ldots,
c'_n)$ and mark the new dead components. It is not difficult to prove
that, continuing in this manner, with probability $1$ we will
eventually generate a finite trajectory $\widehat{\rho}$ and reach a
global state $\vs$ with count vector $(\widehat{c}_1, \widehat{c}_2,
\ldots, \widehat{c}_n)$ such that for each component $i$, either
$\widehat{c}_i \geq k_i$ or $i$ is dead at $\vs$.
% (see \cite{techreport} for the details).  
We then check if $\widehat{\rho}$
satisfies $\varphi$ and update the score associated with the
statistical test described below.

The parameters for the test are $\delta , \alpha, \beta$, where
$\delta$ is the size of the indifference region and $(\alpha, \beta)$
is the strength of the test, with $\alpha$ bounding the Type~I errors
(false positives) and $\beta$ bounding the Type~II errors (false
negatives). These parameters are to be chosen by the user. We generate
finite i.i.d. sample trajectories sequentially. We associate a
Bernoulli random variable $x_{\ell}$ with the sample $\rho_{\ell}$ and
set $x_{\ell} = 1$ if $\rho_{\ell} \in Trj_{\vs^{in}}(\varphi)$ and
set $x_{\ell} = 0$ otherwise. We let $c_m = \sum_{\ell} x_{\ell}$ and
compute the score $SPRT$ via \smallskip

\hspace*{\fill}
$
\displaystyle
SPRT =  \frac{(\gamma^+)^{c_m} (1-\gamma^+)^{n - c_m}}
             {{(\gamma^-)}^{c_m} (1-\gamma^-)^{n - c_m}}
$
\hspace*{\fill}
\smallskip

Here $\gamma^+ = \gamma + \delta$ and $\gamma^- = \gamma - \delta$. If
$SPRT \geq \frac{1-\beta}{\alpha} $, we declare $\cD \models^{trj}
\widehat{P}_{\geq r} \varphi$. If $SPRT \leq \frac{\beta}{1 - \alpha}
$, we declare $\cD \not\models^{trj} \widehat{P}_{\geq \gamma}
\varphi$.  Otherwise, we draw one more sample and repeat.

This test is then extended to handle formulas of the form $\lnot\psi$ and $\psi_1 \vee \psi_2$ in the usual way \cite{jha}.
It is easy to establish the correctness of this statistical model checking procedure.

\section{Experimental results}
\label{section:results}

We have tested our statistical verification procedure on two
probabilistic distributed algorithms: (i)~a leader election protocol
for a unidirectional ring of anonymous processes by Itai and Rodeh
\cite{itairodeh,fokkinkpang} and (ii)~a randomized solution to the dining
philosophers problem by Lehman and Rabin \cite{LR81}. 

For comparison with Markov chain based verification we used the
probabilistic model checking tool \Prism, which tackles these two
examples as case studies \cite{prism,prismweb}. Since \Prism\ does not
currently support SMC for BLTL specifications, we used the
simulation based approximate verification feature of \Prism.  We
compared the time taken by our SMC procedure with that of approximate
verification in \Prism\ for roughly the same number of simulations,.
%TT explained the methodology a bit. Please check I am saying the right thing.
% M Modified sligthly

In the leader election protocol, each process randomly chooses an
identity from $\{1,2,\ldots,N\}$, and passes it on to its neighbour.
If a process receives an identity lower than its own, the message is
dropped. If the identity is higher than its own, the process drops out
of the election and forwards the message. Finally, if the identity is
the same as its own, the process forwards the message, noting the
identity clash.  If an identity clash is recorded, all processes with
the highest identity choose a fresh identity and start another round.

We have built a DMC model of this system in which each process and
channel is an agent.  Messages are transferred between processes and
channels via synchronizations.  For simplicity, all channels in our
implementation have capacity 1.  We can easily construct higher
capacity channels by cascading channels of capacity 1 while staying
within the DMC formalism.

The challenge in modelling the dining philosphers problem as a DMC is
to represent the forks between philosophers, which are typically
modelled as shared variables.  We use a deterministic round robin
protocol to simulate shared variables.  The same technique can be used
for a variety of other randomized distributed algorithms presented as
case studies for \Prism.

\begin{figure}[t]
  \parbox{0.48\textwidth}{
    \includegraphics[width=0.48\textwidth]{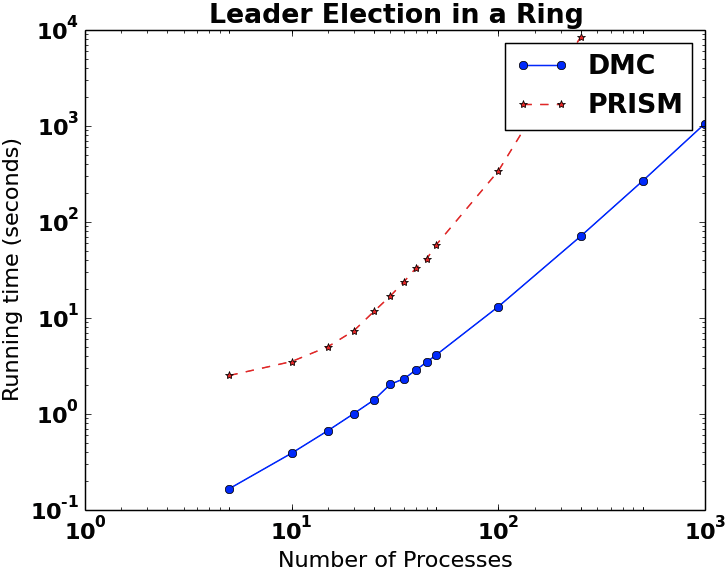}
  }
  \hfill
  \parbox{0.48\textwidth}{
    \includegraphics[width=0.48\textwidth]{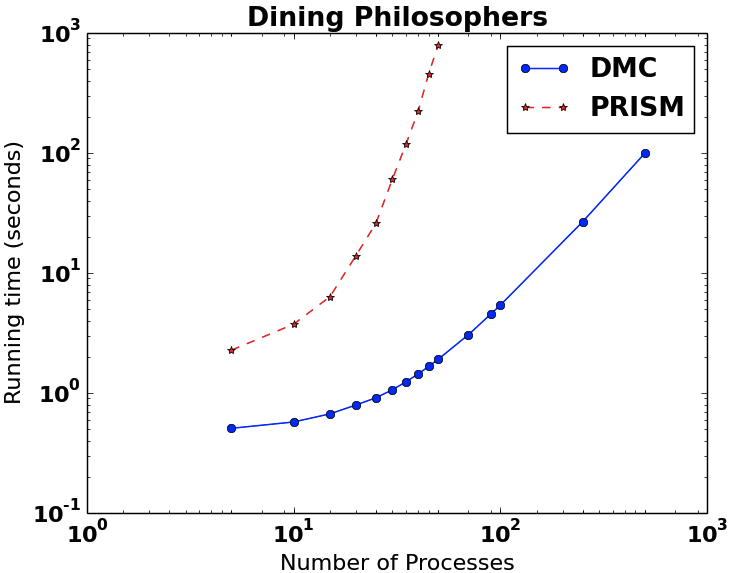}
  }
  \caption{Comparison of simulation times in DMC and \Prism}
  \label{figure:experiment}
\end{figure}

We ran our trajectories based SPRT procedure on a Linux server (Intel
Xeon 2.30~GHz, 16 core, 72GB RAM).  For the first example, we verified
that a leader is elected with probability above 0.99 within $N$
rounds, for a ring of $N$ processes, upto $N=1000$.  For the
dining philosophers, we verified
%T is there a way to desribe this property in one sentence?
that every philosopher eventually eats, upto $N=500$ philosophers.
This property could not be checked using approximate verification in
\Prism\ because it exceeded the simulation bounds even for $N=11$.  To
compare with \Prism, we introduced a boolean
%T could not be checked for N = ?
variable that is set when a philosopher eats for the first time and
verified the property that with probability above 0.95, a fixed
fraction (0.4) of the philosophers have eaten at least once within a
bounded number of steps.

We tested the same properties on the \Prism\ model, using simulation
based approximate verification.  For a fair comparison, we ensured
that the number of simulation runs in \Prism\ were approximately the
same as the average number of simulation runs required for SPRT
verification in the DMC model.

In Fig.~\ref{figure:experiment}, we have compared the running time for
SPRT model-checking in the DMC model against the running time for
approximate verification in \Prism.  The $x$-axis is the number of
processes in the system and the $y$-axis is the running time, in
seconds.  Both axes are rendered in a log scale.  In \Prism, we could
not check the leader election property beyond $N=250$ and, for the
dining philosophers example, the largest system we could run had
$N=50$.  The experiments show that simulations using asynchronous
trajectories are 10 to 100 times faster.

\section{Conclusion}
\label{section:conclusion}

We have formulated a distributed probabilistic system model called
DMCs.  Our model achieves a clean mix of concurrency and probabilistic
dynamics by restricting synchronizations to be deterministic. Our key
technical contribution is the construction of a probability measure
over the $\sigma$-algebra generated by the (interleaved) trajectories
of a DMC. This opens up the possibility of using partial order
reduction techniques to efficiently verify the dynamic properties of a
DMC.  As a first step in this direction we have developed a SPRT based
statistical model checking procedure for the logic
$PBLTL^{\otimes}$. Our experiments suggest that our method can handle
systems of significant sizes.
%T changed slightly

The main partial order concept we have used is to group trajectories
into equivalence classes.  One can also explore how ample
sets~\cite{clarkebook} and related notions can be used to model check
properties specified in logics such as PCTL~\cite{baierbook}. Another
possibility is to see if the notion of finite unfoldings from Petri
net theory can be applied in the setting of DMCs~\cite{McMillan,
  esparzabook}.

In the two examples we have discussed, the specification has a global
character, since it mentions every agent in the system.  In many
specifications, only a few agents may be mentioned.  If the system is
loosely coupled, we can check whether the required property is
fulfilled without having to exercise all the agents. This will lead to
additional computational gains.

One observation is that in standard randomized distributed algorithms,
like the case studies in \cite{prismweb}, probabilistic moves are
always local.  The DMC model allows synchronous probabilistic moves
where the probability distribution is influenced by information
obtained through communication.  This allows us to model situations
such as an actuator for an implanted device making probabilistic
transitions based on information received from external sensors.  In
such situations, the overall dynamics would typically be too complex
to explore exhaustively using traditional model checking techniques,
but statistical model checking can be applied provided we can perform
efficient simulations, which we have demonstrated is possible with
DMCs.

We currently allow agents to gain complete information about the state
of the agents they synchronize with. In practice, only a part of this
state may/should be exposed. An interesting extension would be
point-to-point asynchronous communications through bounded buffers
using a finite message alphabet.

Finally, though almost all the case studies in \cite{prismweb} can be
recast as DMCs, it will be fruitful to understand the theoretical and
practical limitations of deterministic communications in a distributed
probabilistic setting.

\bibliographystyle{plain}

\bibliography{dcmdp}

\begin{thebibliography}{10}

\bibitem{albert1}
S.~Abbes and A.~Benveniste.
\newblock True-concurrency probabilistic models: Branching cells and
  distributed probabilities for event structures.
\newblock {\em Inf. Comput.}, 204(2):231--274, 2006.

\bibitem{albert2}
S.~Abbes and A.~Benveniste.
\newblock True-concurrency probabilistic models: Markov nets and a law of large
  numbers.
\newblock {\em Theor. Comput. Sci.}, 390(2-3):129--170, 2008.

\bibitem{baierbook}
C.~Baier and J.P. Katoen.
\newblock {\em Principles of Model Checking}.
\newblock MIT Press, 2008.

\bibitem{hermanns}
J.~Bogdoll, L.M.F. Fioriti, A.~Hartmanns, and H.~Hermanns.
\newblock Partial order methods for statistical model checking and simulation.
\newblock In {\em FMOODS/FORTE}, pages 59--74, 2011.

\bibitem{clarke}
E.M. Clarke, J.R. Faeder, C.J. Langmead, L.A. Harris, S.K. Jha, and A.~Legay.
\newblock Statistical model checking in {BioLab}: Applications to the automated
  analysis of t-cell receptor signaling pathway.
\newblock In {\em CMSB}, pages 231--250, 2008.

\bibitem{clarkebook}
E.M. Clarke, O.~Grumberg, and D.A. Peled.
\newblock {\em Model Checking}.
\newblock MIT Press, 1999.

\bibitem{david2011time}
A.~David, K.G. Larsen, A.~Legay, M.~Miku{\v{c}}ionis, and Z.~Wang.
\newblock Time for statistical model checking of real-time systems.
\newblock In {\em Computer Aided Verification}, pages 349--355. Springer, 2011.

\bibitem{tracebook}
V.~Diekert and G.~Rozenberg.
\newblock {\em The Book of Traces}.
\newblock World Scientific, 1995.

\bibitem{esparzabook}
J.~Esparza and K.~Heljanko.
\newblock {\em Unfoldings: A Partial-Order Approach to Model Checking}.
\newblock Monographs in theoretical computer science. Springer-Verlag Berlin
  Heidelberg, 2008.

\bibitem{fokkinkpang}
W.~Fokkink and J.~Pang.
\newblock Variations on {Itai-Rodeh} leader election for anonymous rings and
  their analysis in \textsc{Prism}.
\newblock {\em J. of Universal Computer Science}, 12(8):981--1006, 2006.

\bibitem{baier_survey}
M.~Gr{\"o}{\ss}er and C.~Baier.
\newblock Partial order reduction for {Markov Decision Processes}: A survey.
\newblock In {\em FMCO}, pages 408--427, 2005.

\bibitem{smolka1}
R.~Grosu, X.~Huang, S.A. Smolka, W.~Tan, and S.~Tripakis.
\newblock Deep random search for efficient model checking of timed automata.
\newblock In Fabrice Kordon and Oleg Sokolsky, editors, {\em Monterey
  Workshop}, volume 4888 of {\em Lecture Notes in Computer Science}, pages
  111--124. Springer, 2006.

\bibitem{prism}
A.~Hinton, M.~Kwiatkowska, G.~Norman, and D.~Parker.
\newblock \textsc{Prism}: A tool for automatic verification of probabilistic
  systems.
\newblock In {\em 12th TACAS}, pages 441--444, 2006.

\bibitem{itairodeh}
A.~Itai and M.~Rodeh.
\newblock Symmetry breaking in distributed networks.
\newblock {\em Information and Computation}, 88(1):60--87, 1990.

\bibitem{legay1}
C.~Jegourel, A.~Legay, and S.~Sedwards.
\newblock A platform for high performance statistical model checking--{PLASMA}.
\newblock {\em Tools and Algorithms for the Construction and Analysis of
  Systems}, pages 498--503, 2012.

\bibitem{jha2009statistical}
S.~Jha.
\newblock Statistical analysis of privacy and anonymity guarantees in
  randomized security protocol implementations.
\newblock {\em arXiv preprint arXiv:0906.5110}, 2009.

\bibitem{jha}
S.K. Jha, E.M. Clarke, C.J. Langmead, A.~Legay, A.~Platzer, and P.~Zuliani.
\newblock A {Bayesian} approach to model checking biological systems.
\newblock In {\em CMSB}, pages 218--234, 2009.

\bibitem{DBLP:conf/wabi/LangmeadJ07}
C.J Langmead and S.K. Jha.
\newblock Predicting protein folding kinetics via temporal logic model
  checking.
\newblock In {\em WABI}, pages 252--264, 2007.

\bibitem{viswanathan1}
A.~Legay and M.~Viswanathan.
\newblock Simulation + hypothesis testing for model checking of probabilistic
  systems.
\newblock In {\em QEST}, page~3. IEEE Computer Society, 2009.

\bibitem{LR81}
D.~Lehmann and M.~Rabin.
\newblock On the advantage of free choice: {A} symmetric and fully distributed
  solution to the dining philosophers problem (extended abstract).
\newblock In {\em Proc. 8th Annual ACM Symposium on Principles of Programming
  Languages (POPL'81)}, pages 133--138, 1981.

\bibitem{McMillan}
K.L. McMillan.
\newblock A technique of state space search based on unfolding.
\newblock {\em Formal Methods in System Design}, 6(1):45--65, 1995.

\bibitem{DBLP:journals/mst/JesiPS96}
G.~Pighizzini S.~Jesi and N.~Sabadini.
\newblock Probabilistic asynchronous automata.
\newblock {\em Mathematical Systems Theory}, 29(1):5--31, 1996.

\bibitem{winskel}
D.~Varacca, H.~V{\"o}lzer, and G.~Winskel.
\newblock Probabilistic event structures and domains.
\newblock {\em Theor. Comput. Sci.}, 358(2-3):173--199, 2006.

\bibitem{prismweb}
Various.
\newblock \textsc{Prism}\ case studies.
\newblock \url{http://www.prismmodelchecker.org/casestudies}, 2013.

\bibitem{Wald:1947}
A.~Wald.
\newblock {\em Sequential Analysis}.
\newblock John Wiley and Sons, 1st edition, 1947.

\bibitem{younesthesis}
H.L.S. Younes.
\newblock {\em Verification and planning for stochastic processes with
  asynchronous events}.
\newblock PhD thesis, Pittsburgh, PA, USA, 2004.
\newblock AAI3159989.

\bibitem{younes}
H.L.S. Younes and R.G. Simmons.
\newblock Probabilistic verification of discrete event systems using acceptance
  sampling.
\newblock In {\em CAV}, pages 223--235, 2002.

\bibitem{DBLP:journals/ita/Zielonka87}
W.~Zielonka.
\newblock Notes on finite asynchronous automata.
\newblock {\em ITA}, 21(2):99--135, 1987.

\end{thebibliography}

\end{document}